\documentclass[%
 aip,
 amsmath,amssymb,
 reprint,%
]{revtex4-2}

\usepackage{graphicx}
\usepackage{dcolumn}
\usepackage{bm}

\usepackage[utf8]{inputenc}
\usepackage[T1]{fontenc}
\usepackage{mathptmx}
\usepackage{etoolbox}
\usepackage{xcolor}
\usepackage{ulem}

\makeatletter
\def\@email#1#2{%
 \endgroup
 \patchcmd{\titleblock@produce}
  {\frontmatter@RRAPformat}
  {\frontmatter@RRAPformat{\produce@RRAP{*#1\href{mailto:#2}{#2}}}\frontmatter@RRAPformat}
  {}{}
}%
\makeatother
\begin{document}

\preprint{AIP/123-QED}

\title{Shot noise generated by subpopulations of neural networks}
\author{S. Yu. Kirillov}
 \email{skirillov@ipfran.ru}
 \affiliation{A.V. Gaponov-Grekhov Institute of Applied Physics of the Russian Academy of Sciences, Ulyanova Street 46, Nizhny Novgorod 603950, Russia
}%
\author{O. A. Goryunov}
 \email{ogoryunov@hse.ru}
 \affiliation{A.V. Gaponov-Grekhov Institute of Applied Physics of the Russian Academy of Sciences, Ulyanova Street 46, Nizhny Novgorod 603950, Russia}
  \affiliation{National Research University Higher School of Economics, 25/12 Bol’shaya Pecherskaya Street, Nizhny Novgorod 603155, Russia
}%
\author{J. Zhu}
 \email{jinjiezhu@nuaa.edu.cn}
 \affiliation{State Key Laboratory of Mechanics and Control for Aerospace Structures, College of Aerospace Engineering, Nanjing University of Aeronautics and Astronautics, Nanjing 210016, China
}%
\author{V. V. Klinshov}%
 \email{vklinshov@hse.ru}
  \affiliation{A.V. Gaponov-Grekhov Institute of Applied Physics of the Russian Academy of Sciences, Ulyanova Street 46, Nizhny Novgorod 603950, Russia
}%
 \affiliation{National Research University Higher School of Economics, 25/12 Bol’shaya Pecherskaya Street, Nizhny Novgorod 603155, Russia
}%



\newcommand{\corr}[2]{#2}

\begin{abstract}
While recent advances in next-generation neural mass models provide exact descriptions of densely coupled neural populations in the thermodynamic limit, populations in vivo remain strictly finite in size. Finite-size effects introduce stochastic fluctuations whose impact on network dynamics depends on their spectral content. Furthermore, coupling between different populations is typically sparse, meaning that only a small, random subset of neurons from one population projects connections to another. This subset (a subpopulation) produces an output signal that is inherently noisy. Given that the subpopulation constitutes only a fraction of the full population, its shot noise differs from that of the whole population in both intensity and spectral shape. In the present work, we analyze these differences and demonstrate that they depend non-trivially on subpopulation size. Using a generalization of our nesting method, we derive an analytical expression for the power spectral density of subpopulation shot noise, which shows excellent agreement with direct numerical simulations. Unlike many previous studies that rely on mathematically convenient but unrealistic Lorentzian distributions (whith diverging moments), our approach accounts for more realistic, non-Lorentzian distributions of local neuron parameters using a previously developed reduction technique. These results provide a foundation for a new class of stochastic mean-field models for hierarchical neural networks. Such models can now incorporate the correct, size-dependent frequency spectrum of subpopulation shot noise. Crucially, this spectrum is not a simple scaled version of the full population’s noise. instead, it arises from a non-trivial mixture of two distinct spectral components. This is essential for networks with dense local connectivity and sparse inter-population connectivity.
\end{abstract}

\maketitle

\begin{quotation}
The brain is composed of many local groups of neurons that are densely connected internally but only sparsely connected to one another. To understand how such large-scale networks operate, it is common to use the so-called neural mass models, which describe the average activity of a whole population rather than every single neuron. However, when one population sends signals to another, it does so only through a small, random subset of its neurons — a ``subpopulation''. Because these subpopulations are small, their signals are inherently noisy. In this study, we show that this kind of noise, known as shot noise, has a fundamentally different frequency spectrum than the noise produced by the full population. We derive an analytical formula for this spectrum, revealing how its properties change with subpopulation size. Our work provides the first analytical framework for modelling the spectral fingerprint of noisy signals from subpopulations, paving the way for a new generation of stochastic brain network models that respect the sparse, hierarchical connectivity found in real neural systems.
\end{quotation}

\section{Introduction}

One of the key features of large-scale networks of biological neurons is their modular architecture: neurons form local populations with relatively high connection density, whereas inter-population connections remain sparse \cite{meunier2010modular,Martijn2017hubs,Kim2020sw}. This allows us to model the brain as a network of interacting populations. To describe the dynamics within individual populations, researchers often employ neural mass models. These are reduced dynamical systems that capture the average activity (such as the mean firing rate or membrane potential) of a large population of interacting neurons. Their main advantage is that they dramatically reduce computational complexity while preserving the key collective dynamics, thus facilitating efficient numerical simulation and analytical tractability \cite{wilson1972excitatory,jansen1995electroencephalogram,deco2008dynamic,schwalger2017towards}.

Recently, the so-called next-generation of neural mass models has been developed \cite{montbrio2015macroscopic}. Unlike traditional phenomenological models, these are derived mathematically from the microscopic dynamics of neurons and become exact in the thermodynamic limit (i.e., for an infinite population size). This framework has been extended to sparse neural networks \cite{divolo2018transition,goldobin2021reduction}, to populations with adaptation and plasticity \cite{gast2021mean,chen2022exact,eydam2024control,pietras2025low,fennelly2025mean}, and to balanced excitatory-inhibitory networks \cite{bi2021asynchronous,Goldobin2025syn}.

However, real neural populations are finite, leading to stochastic fluctuations around the thermodynamic (infinite-size) behavior  — a phenomenon known as finite-size effects. These effects cannot be ignored when comparing model predictions to experimental recordings or when simulating large-scale network dynamics \cite{brunel1999fast,dumont2017stochastic}.

In our recent work, we developed a technique to describe finite-size effects by incorporating stochastic terms into neural mass models \cite{klinshov2022shot}. We derived analytical expressions for the power spectrum of these stochastic fluctuations, which we termed shot noise, and showed that the spectrum can exhibit non-trivial spectral peaks. These peaks may lead to resonance effects, potentially altering the collective dynamics of neural networks. The interplay between noise, finite size, and network dynamics has been actively investigated, including studies of the role of connectivity disorder in finite-size networks \cite{Greven2026rnd},  the effects of asymmetric Cauchy noise \cite{Ageeva2025asymm} and the emergence of brain rhythms due to finite size fluctuations \cite{Nandi2024gamma}.

When modeling large-scale networks consisting of many coupled populations, each population can be described by a neural mass model. However, accurately modeling the connections between these populations requires accounting for the sparsity and randomness of inter-population connections. Consequently, each neuron in one population receives input not from all neurons of another population, but only from a small, randomly selected subset — a subpopulation. This structural feature significantly increases the potential role of shot noise in network dynamics. The signal transmitted through such sparse connections is inherently noisy, and the strength of this noise increases as the subpopulation size decreases. Therefore, understanding how the power spectrum of subpopulation shot noise depends on subpopulation size is essential for accurate modeling of sparse inter-population communication in hierarchical brain networks.

The present work is devoted to the study of shot noise generated by subpopulations within densely coupled neural populations. We show that the shot noise of subpopulations differs from that of the full population not only in intensity but also in the shape of the spectrum, with its properties depending non-trivially on subpopulation size. Using a generalization of our nesting method, we derive an analytical expression for the power spectral density of this signal, which demonstrates excellent qualitative and quantitative agreement with direct numerical simulations.

It is important to note that, unlike most studies which rely on a Lorentzian distribution of neuron parameters for mathematical convenience, we adopt a more realistic distribution. This choice avoids unphysical features associated with the Lorentzian while remaining analytically tractable. The importance of going beyond the Cauchy distribution has been increasingly recognized, with recent work developing mean-field reductions for rational approximations of Gaussian and uniform distributions, as well as for bimodal and q-Gaussian heterogeneity \cite{klinshov2021reduction,pyragas2021dynamics,pyragas2022mean,pyragas2023mean,pyragas2024mean}. This question is part of a broader investigation into how neural heterogeneity — whether in spike thresholds, excitability, or connectivity — shapes network dynamics and computation \cite{gast2024neural}.

To describe the macroscopic dynamics of our network, we use a reduction method we developed earlier \cite{klinshov2021reduction}. To derive the power spectrum of shot noise for its subpopulations, we employ a generalization of our previously proposed nesting method \cite{klinshov2022shot}. By combining the concepts of shot noise and neural mass models — extending our earlier analysis of whole populations to the case of subpopulations — we provide, to the best of our knowledge, the first analytical framework for modeling the spectral properties of subpopulation shot noise with realistic heterogeneity. These results open the path to a new class of stochastic mean-field models for hierarchical neural networks with sparse inter-population connections.

\section{Microscopic model of the population}

Consider a population of globally coupled quadratic integrate-and-fire neurons of large, but finite size $N$:
\begin{equation}\label{a1_01}
   \dot{V_j}=V_j^2+\eta_j+I_{inp}(t),
\end{equation}
where  $V_j$ is the membrane potential of the $j$-th neuron, $j=1,\ldots,N$, $\eta_j$ are the individual bias currents determining the local dynamics, and $I_{inp}(t)$ is the common input to all the neurons consisting of the recurrent and external parts:
\begin{equation}\label{eq:I_inp}
     I_{inp}(t)=I(t) + J s(t),
\end{equation}
where $I(t)$ is the external input current, $J$ is the coupling strength inside the population, and  $s(t)$ is the recurrent synaptic current defined as
\begin{equation}\label{a1_02}
s(t)=\frac{1}{N}\sum_{j=1}^N\sum_{k \backslash t_j^k<t}\int_{-\infty}^t dt' a_{\tau}(t-t')\delta(t'-t_j^k).
\end{equation}
Here $t_j^k$ is the time of generation of the $k$-th spike by the $j$-th neuron, $\delta(t)$ is the Dirac delta function, $a_{\tau}(t)=e^{-t/\tau}/\tau$ is the activation function of a single synapse, where $\tau$ is the characteristic decay time of the synaptic current. In what follows, we take $\tau\to 0$, which corresponds to $a_\tau(t)=\delta(t)$.

The signal $s(t)$ represents the output of the entire population. We will also be interested in the output signals of its subpopulations which we will define later in Sec.~\ref{sec:subpop}.

The individual parameters $\eta_j$ are drawn from some continuous density distribution $g(\eta)$. It is common to choose a Lorentzian distribution for $g(\eta)$, which simplifies theoretical analysis in many cases \cite{montbrio2015macroscopic}. However, this choice is not without drawbacks. Its heavy tails lead to undefined higher-order moments and are physically implausible, implying an unrealistically high likelihood of extreme parameter values for individual neurons. In the present study we adopt the following distribution:
\begin{equation}\label{eq:g}
g(\eta)=\frac{\sqrt{2}}{\pi}\frac{1}{\Delta^4+(\eta-\zeta)^4}.
\end{equation}
This is a symmetric rational distribution centered at $\zeta$ with a half-width at half-maximum equal to $\Delta=1$. Unlike the Lorentzian distribution, its tails decay as $\sim \eta^{-4}$, ensuring finite moments of all orders. This property not only improves the physical plausibility of the model but also enables the use of larger integration steps in numerical simulations, significantly reducing computational costs.

For finite-size populations, the collective dynamics may depend not only on the overall form of the underlying parameter distribution $g(\eta)$, but also on the specific  sampling method used to select the individual parameter values \cite{kirillov2025collective}. Here, we draw individual parameters $\eta_j$ from the distribution \eqref{eq:g} randomly and independently by inverse transform sampling (see Appendix A for details).

\section{Neural mass model}

We first consider the thermodynamic limit $N\to\infty$. In this limit, the state of the system can be described by the distribution function $\rho(V|\eta,t)$, and the output signal $s(t)$ coincides with the average firing rate of the entire population $r(t)$ (see Appendix B for details). For clarity, we will use $s(t)$ to denote the output signal of a finite-size population and $r(t)$ for the output of an infinite population.

As demonstrated in Ref. \cite{montbrio2015macroscopic}, for a Lorentzian distribution $g(\eta)$ the population dynamics can be described by a reduced two-dimensional neural mass model for the mean firing rate $r$ and the mean membrane potential $v$. Building on this result, we generalized the approach to a broader class of parameter distributions \cite{klinshov2021reduction}. Specifically, we proved that if $g(\eta)$ is a rational function with $n$ simple poles lying in the lower complex half-plane, then in the thermodynamic limit the population dynamics can be described by a system of $n$ ODEs for the complex order parameters $w_k$:
\begin{equation}\label{eq:16}
\dot{w}_k=i\big[\bar{\eta}_k+Jr-w_k^2+I(t)\big],\quad k=1,\dots,n.
\end{equation}
Here, $\bar{\eta}_k$ are the poles of $g(\eta)$ with $\operatorname{Im}\bar{\eta}_k<0$, and the mean firing rate $r$ is given by a linear combination of the $w_k$:
\begin{equation}\label{eq:15}
r=\Re\sum_{k=1}^n a_k w_k,
\end{equation}
where $a_k=2i\big[g(\bar{\eta}_k)\big]^2/g'(\bar{\eta}_k)$ are coefficients determined by the residues of $g(\eta)$ at its poles.

For $n\geq 2$, the order parameters $w_k$ lack any obvious physical interpretation on their own. Nevertheless, they provide a closed description of the population dynamics.  In the specific case $n=2$, when $g(\eta)$ is defined by \eqref{eq:g}, the resulting neural mass model can be described as two complex equations for $w_1$ and $w_2$. Equivalently it can be expressed as four real equations for the variables $p_k=\Re(w_k)$, $q_k=\Im(w_k)$, where $k=1,2$:
\begin{equation}\label{a3_11}
\begin{aligned}
&\dot{p}_1=\frac{\sqrt{2}}{2}+2 p_1 q_1\\
&\dot{q}_1=\zeta+\frac{\sqrt{2}}{2}+J r - p_1^2+q_1^2+I(t)\\
&\dot{p}_2=\frac{\sqrt{2}}{2}+2 p_2 q_2\\
&\dot{q}_2=\zeta-\frac{\sqrt{2}}{2}+J r - p_2^2+q_2^2+I(t)
\end{aligned}
\end{equation}
The firing rate in this representation is given by:
\begin{equation}
    r=\frac{1}{2\pi}\Big\{p_1-q_1+p_2+q_2\Big\}.
\end{equation}

The detailed derivation of this neural mass model and its bifurcation analysis can be found in Appendix B. In the autonomous case ($I(t)\equiv 0$), numerical bifurcation analysis of \eqref{a3_11} reveals that the system can exhibit one or two stable equilibrium states depending on the parameters $J$ and $\zeta$, with no limit cycles observed. In what follows, we set the parameters such that the model \eqref{a3_11} possesses a unique stable equilibrium state.

\section{Shot noise of the entire population}

For infinite $N$, the output $r(t)$ of the population is given by the neural mass model \eqref{a3_11} and is therefore constant in the absence of external input: $r=r_0$. However, when the population size is finite, it can be described by the neural mass model only approximately, and its output $s(t)$ exhibits fluctuations near the neural mass  model's the steady state $r_0$. In order to study these finite-size fluctuations, we employ the ``nested'' setting introduced in \cite{klinshov2022shot}. In this approach, a finite-size population of size $N$ is viewed as a part of a much larger ``superpopulation'' of size $N_+\to\infty$, which shares the same asymptotic parameter distribution $g(\eta)$. The infinite size of the superpopulation allows its dynamics to be described \emph{exactly} by the neural mass model. The finite-size subsystem, however, does not replicate this behavior: its signal deviates from the superpopulation's output. To characterize this deviations (fluctuations) in a normalized form (essential for comparing results across different $N$), we introduce the \textit{free} shot noise:
\begin{equation}
\tilde{\chi}(t) = \sqrt{N} \big(s(t) - r(t)\big),
\end{equation}
defined as the normalized difference between the output of the finite population and that of the infinite superpopulation, under the condition that both receive \emph{the same external input} $I_{inp}(t)$.

\begin{figure*}
	\includegraphics[width=0.9\textwidth]{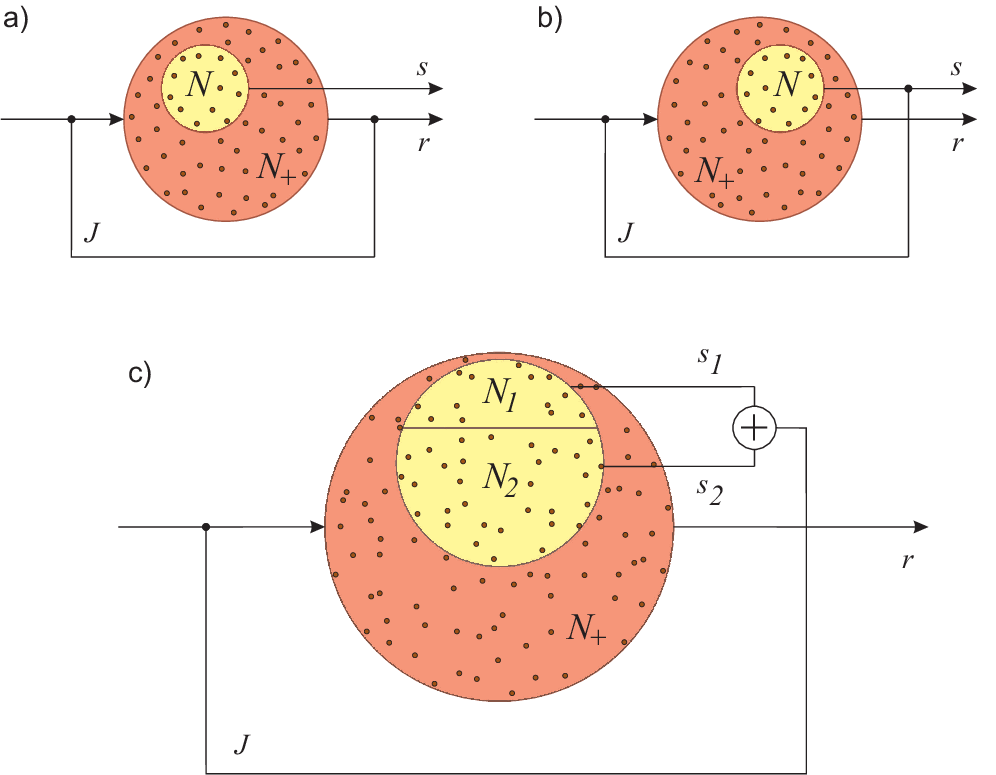}
	\caption{Three versions of the nestled setting used in the analysis. The finite population of the size $N$ is always a part of the infinite superpopulation of the size $N^+\to\infty$, and all the neurons allways receive the same input. (a)  All the neurons receive the input from the infinite superpopulation. (b) All the neurons receive the input from the finite population. (c) The finite population isw divided into two subpopulations, all the neurons receive a combined input from the both subpopulations.}\label{fig:Nested_Sub}
\end{figure*}

Next, we characterize the spectral properties of the fluctuations. We first compute the power spectral density of the free shot noise. For this, we assume that all neurons of the superpopulation (including those belonging to the embedded finite population) receive input from the \emph{entire superpopulation}, which implies that this input is constant in time (see Fig.~\ref{fig:Nested_Sub}(a)). Under this assumption, each $j$-th neuron in the finite population receives a constant input $I_j=\eta_j+Jr_0$ and consequently generates an output spike train that is a Dirac comb with individual frequency $\nu_j=\sqrt{\eta_j+Jr_0}/\pi$. The total output of the finite population can thus be written as a sum of such Dirac combs, and its power spectrum is given by
\begin{equation}\label{eq:W0}
	\tilde{W}(\nu) = \sum_{q=1}^{\infty}\frac{\nu^2}{q^3}\; g_\nu\left(\frac{\nu}{q}\right),
\end{equation}
where $g_\nu(\nu)$ is the frequency distribution density. This density can be obtained from the parameter distribution $g(\eta)$ via $g_\nu(\nu)=g(\eta)|d\eta/d\nu|$. For the specific distribution \eqref{eq:g} used in this work,
\begin{equation}
	g_\nu(\nu) =\frac{2\sqrt{2}\pi\nu}{1+\big[(\pi\nu)^2-Jr_0-\zeta\big]^4}.
\end{equation}
The detailed derivation of the power spectrum of free shot noise is provided in Appendix C.

We now proceed to the second step and consider the nested setting in which all neurons receive input from the \emph{finite population} (see Fig.~\ref{fig:Nested_Sub}(b)). In this configuration, the dynamics of the finite population remain closed, as it receives input only from itself and does not depend on the other neurons of the superpopulation. The superpopulation is retained solely for theoretical analysis, as it still can be described exactly by the neural mass model almost identical to \eqref{a3_11}. The key modification in this setting is that the superpopulation receives recurrent input $Js$ rather than $Jr$. Since $s=r+\tilde{\chi}/\sqrt{N}$ by the definition of free shot noise, the superpopulation can be described by \eqref{a3_11} with an effective external input $I(t)=J\tilde{\chi}/\sqrt{N}$. For large $N$, the shot noise term is small, and the total input to individual neurons remains close to constant.

We now examine in more detail the properties of signals $s(t)$ and $r(t)$. The high-frequency stochastic term $\tilde{\chi}$ causes low-frequency macroscopic fluctuations\cite{klinshov2022shot} of magnitude $\sim1/\sqrt{N}$ in the neural mass model. As a result, its output $r$ fluctuates near the steady state $r_0$:
\begin{equation}
    r(t)=r_0+\psi(t)/\sqrt{N},
\end{equation}
where $\psi(t)$ represents the normalized macroscopic fluctuations. To determine $\psi(t)$, we linearize system \eqref{a3_11} near its steady state and compute its impulse response $h(t)$. This yields:
\begin{equation}\label{eq:macro}
    \psi(t)=Jh(t)\ast  \tilde{\chi}(t),
\end{equation}
where the asterisk denotes convolution. The output signal of the finite population can then be found as:
\begin{equation}
    s(t)=r(t)+\tilde{\chi}(t)/\sqrt{N}=r_0+(\tilde{\chi}(t)+\psi(t))/\sqrt{N},
\end{equation}
This allows us to define the (normalized) \emph{full shot noise} as the sum of microscale and macroscale components:
\begin{equation}
\chi(t)=\tilde{\chi}(t)+\psi(t)=\tilde{\chi}(t)+Jh(t)\ast \tilde{\chi}(t),
\end{equation}
and to obtain the power spectral density $W(\nu)$ of the full shot noise:
\begin{equation}\label{eq:WJ}
	W(\nu)=\left|1+JS(\nu)\right|^2 \tilde{W}(\nu),
\end{equation}
where $S(\nu)$ is the frequency response (Fourier transform of the impulse response $h(t)$) of the neural mass model. The detailed derivation of $S(\nu)$ is provided in Appendix D.

\section{Shot noise of a subpopulation}\label{sec:subpop}

Now let us study shot noise generated not by the entire population of the size $N$, but by its subpopulation of a certain size. We stay within the framework  of  the ``nested'' setting:  the finite-size population is still a part of the infinite superpopulation, and all the neurons from the superpopulation receive the input from the \emph{finite population}. Let us now divide the finite population into two parts, one of the size $N_1=\alpha N$ and the remaining part of the size $N_2=N-N_1=(1-\alpha)N$ (see Fig.~\ref{fig:Nested_Sub}(c)). The neurons are divided into two subpopulations randomly, so that they both have the same distribution of the bias currents $\eta$ and differ only in size. Let us denote the outputs of the first and the secon subpopulations as $s_1$ and $s_2$, respectively:
\begin{equation}\label{a1_02}
s_l(t)=\frac{1}{N_l}\sum_{j\in P_l}\sum_{k \backslash t_j^k<t}\int_{-\infty}^t dt' a_{\tau}(t-t')\delta(t'-t_j^k),\;l=1,2,
\end{equation}
where $P_l$ is the set of neurons from the $l$-th subpopulation. Note that the signal from each population is normalized over its  size, not the size of the whole population.

Since both subpopulations receive identical input to the superpopulation (from the full finite population), their outputs can be expressed as
\begin{equation}
    s_l=r+\frac{1}{\sqrt{N_l}}\tilde{\chi_l},
\end{equation}
where $\tilde{\chi_l}$ is the normalized \emph{free} shot noise of the $l$-th subpopulation. This allows us to rewrite the macroscopic fluctuations of Eq.~\eqref{eq:macro} in terms of the noise components as:
\begin{equation}
    \psi=Jh\ast \left(\sqrt{\alpha} \tilde{\chi_1}(t) + \sqrt{1-\alpha}\tilde{\chi_2}(t)\right)
\end{equation}
and express  the full shot noise of the first subpopulation as
\begin{equation}
    \chi_{1}=\sqrt{N_1}(s_1-r_0)=\sqrt{\alpha}Jh\ast \left(\sqrt{\alpha} \tilde{\chi_{1}}(t) + \sqrt{1-\alpha}\tilde{\chi_{2}}(t)\right)+\tilde{\chi_{1}}.
\end{equation}
Moving to Fourier space leads to the following equation for the Fourier spectrum of the full shot noise:
\begin{equation}    \phi_1=\left(1+\alpha J S(\nu)\right)\tilde{\phi_{1}}+\sqrt{\alpha(1-\alpha)}JS(\nu)\tilde{\phi_{2}},
\end{equation}
where $\tilde{\phi_{l}}$ is the Fourier spectrum of the free shot noise of the $l$-th subpopulation. The normalized power spectrum of the full shot noise of the first subpopulation $W_\alpha=|\phi_1|^2$ then can be found as
\begin{eqnarray}
    W_\alpha&=&|\tilde{\phi_{1}}|^2|1+\alpha J S|^2+ \alpha(1-\alpha)|JS|^2|\tilde{\phi_{2}}|^2.
\end{eqnarray}
where superscript asterisk represents the complex conjugate. Since the phases of the two spectra $\tilde{\phi_{1}}$ and $\tilde{\phi_{2}}$ are random and independent, the terms including the  products $\tilde{\phi_{1}}\tilde{\phi_{2}}^*$ and $\tilde{\phi_{2}}\tilde{\phi_{1}}^*$ ara zero on average over any finite frequency band and therefore can be ommitted. Taking into account that the normalized power spectra of free shot noises of the both subpopulations are equal and given by \eqref{eq:W0}, we finally obtain
\begin{equation}\label{eq:Wa}
    W_\alpha=\alpha W + (1-\alpha) \tilde{W}.
\end{equation}

\section{Numerical results}

\subsection{Parameters of simulations}
Below are the results of direct numerical simulation for a finite-size network with $N=10^4$, $J=-10$, $\zeta=9.7326$. The obtained data reproduce the individual characteristics of a single realization of such a network, allowing for a quantitative comparison with the predicted envelope.

The system (\ref{a1_01}) was numerically integrated using the Euler method over a time interval of $T=5\times 10^4$ with a step size of $\Delta t=1\cdot 10^{-3}$. The initial phases were chosen randomly and independently from a uniform distribution on the half-interval $\varphi\in[0;2\pi)$ on the Ott–Antonsen manifold. The power spectrum of fluctuations of the mean frequency of collective oscillations in the network (\ref{a1_01}) was computed using the fast Fourier transform with a Hanning window. To reduce the variance of the spectral estimate, Welch’s method was applied: the data were divided into segments of length $2^{22}$ time samples, and the periodograms were averaged over all segments. As an additional step, the data are averaged using a sliding frequency window of width $\Delta\omega=0.012$, which serves to smooth the spectral estimates.

\subsection{Full population spectrum}
\begin{figure*}
	\includegraphics[width=1.0\textwidth]{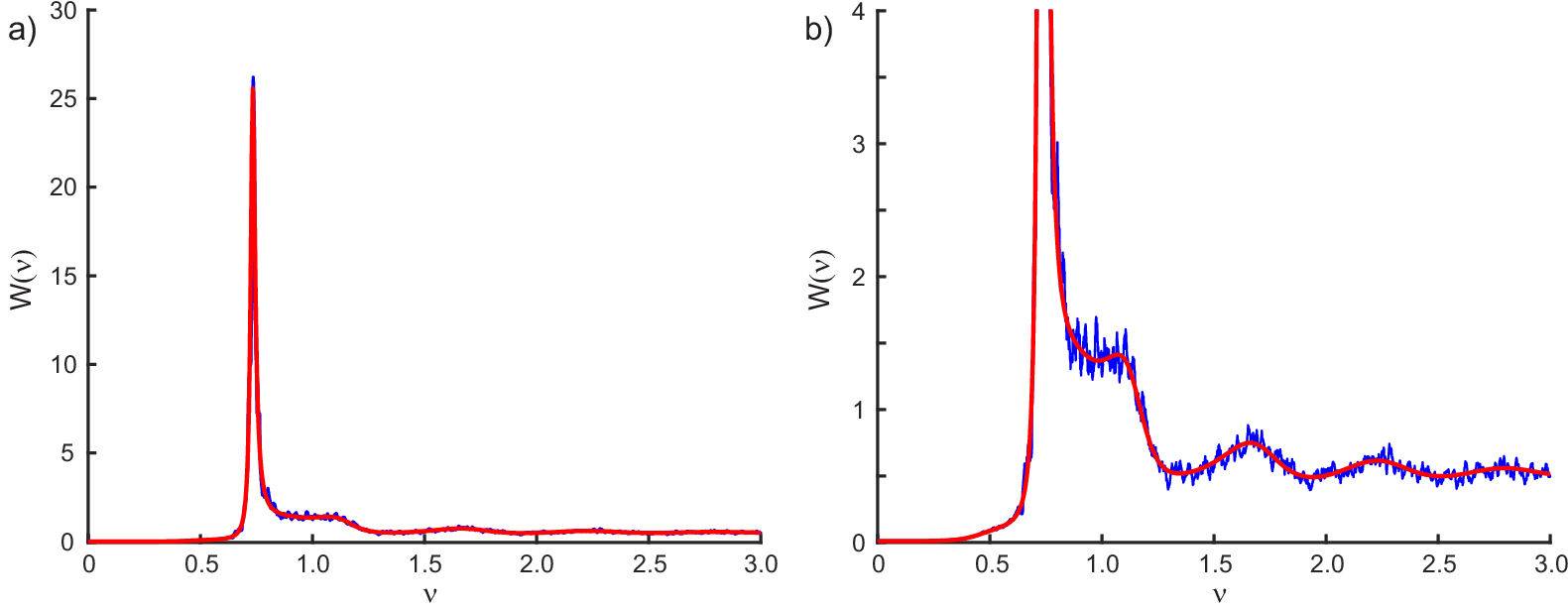}
	\caption{(a) Normalized power spectrum of shot noise in a fully connected network and (b) its magnified segment. The blue line represents the results of numerical simulations, while the red line corresponds to the analytical prediction \eqref{eq:WJ}. Parameters: $N=10^4$, $J=-10$, $\zeta=9.7326$.}\label{fig:full_coupled}
\end{figure*}

Figure~\ref{fig:full_coupled} shows the shot noise power spectrum for a fully connected network.
The spectrum is characterized by a pronounced low-frequency maximum that dominates over the other spectral harmonics. An enlarged fragment of the spectrum (Fig.~\ref{fig:full_coupled}b) reveals a sequence of regularly spaced local maxima, whose height gradually decreases with increasing frequency.

The nature of the shot noise in a fully connected network and the origin of its oscillatory spectrum have been discussed in detail in Ref.~\cite{klinshov2022shot}. Here we briefly note the key features. The frequency of the main peak is determined by the properties of the equilibrium state of the mean-field system, which is a focus. This frequency sets the position of the main maximum in the spectrum. The positions of the remaining peaks are related to the mean frequency of the population: the frequencies of the local maxima are multiples of the mean frequency. Note that the spectrum contains peaks only at higher harmonics of the mean frequency, while at the mean frequency itself there is no pronounced local maximum.

\subsection{Subpopulation spectra}
\begin{figure*}
	\includegraphics[width=1.0\textwidth]{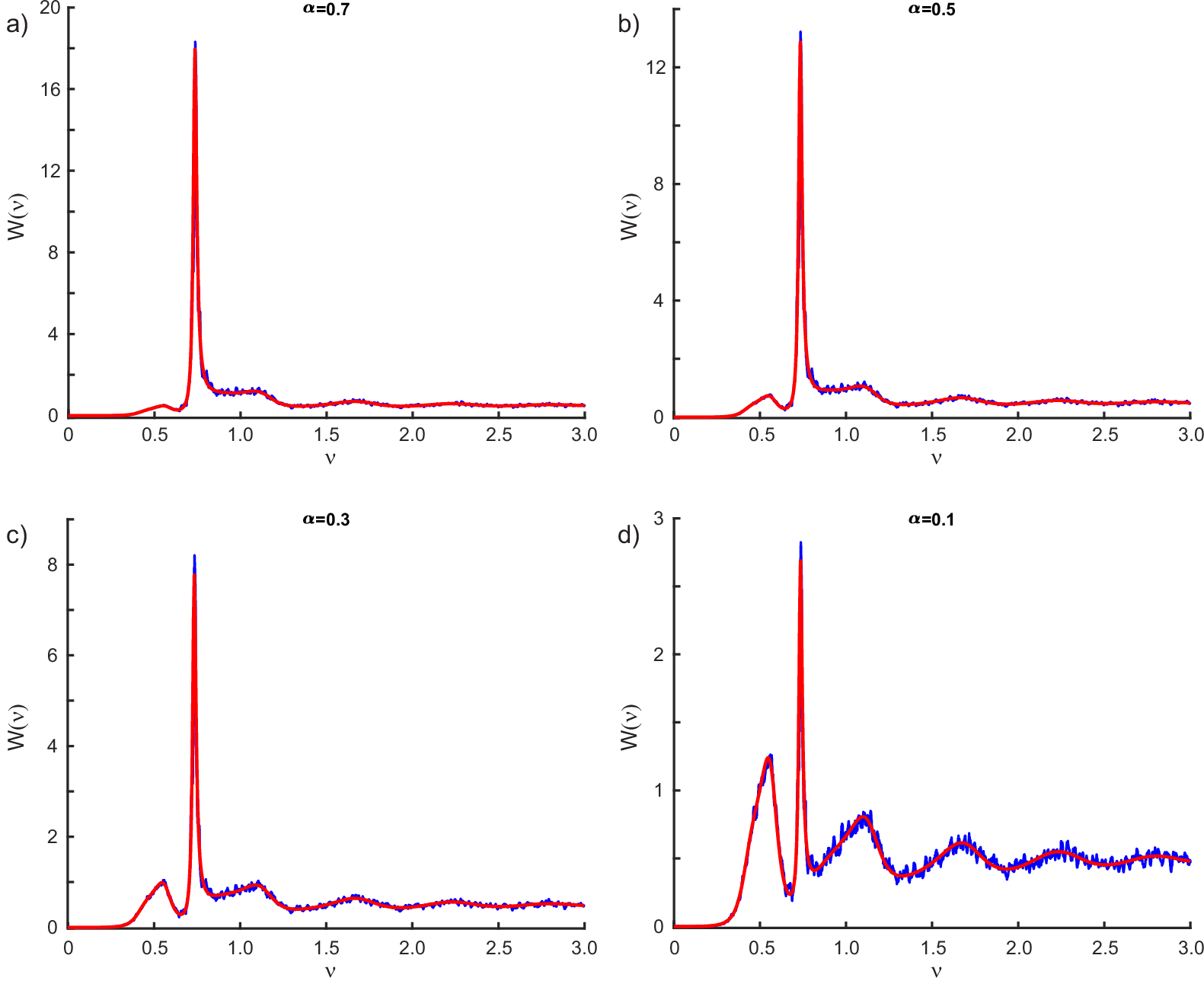}
	\caption{Normalized shot noise spectrum at the output of subpopulations of various sizes in a fully connected network. Panels: (a) $\alpha=0.7$, (b) $\alpha=0.5$, (c) $\alpha=0.3$, (d)  $\alpha=0.1$. Parameters: $N=10^4$, $J=-10$, $\zeta=9.7326$.}\label{fig:spctr_alpha}
\end{figure*}

Now consider the properties of shot noise at the output of individual subpopulations of a fully connected network. The elements of each subpopulation were randomly selected from the full network. The results of the study are shown in Fig.~\ref{fig:spctr_alpha}.

For $\alpha=1.0$ (the subpopulation exactly matches the full network), the subpopulation spectrum is identical to that shown in Fig.~\ref{fig:full_coupled}. As the fraction of network elements in the subpopulation decreases ($\alpha<1$), systematic changes in the spectrum are observed, most pronounced in the low-frequency region. At $\alpha=0.7$ (Fig.~\ref{fig:spctr_alpha}a), the amplitude of the main peak at the focus frequency decreases, while a local maximum emerges near the first harmonic of the mean frequency. A slight decrease in spectral power is also noticeable near the second harmonic of the mean frequency. In the high-frequency region, the spectrum remains almost unchanged.

With further reduction of $\alpha$ ($\alpha=0.5$ and $\alpha=0.3$, Fig.~\ref{fig:spctr_alpha}b,c), the trend of decreasing amplitude of the peak at the focus frequency continues, and the spectral power near the first harmonic of the mean frequency keeps growing.

In the case of small subpopulations ($\alpha=0.1$, Fig.~\ref{fig:spctr_alpha}d), the shot noise spectrum approaches that of free shot noise (characteristic of independent oscillators). A sharp peak at the focus frequency persists, indicating that the influence of the mean-field dynamics of the full network remains even when the subpopulation includes only a small fraction of elements.

The analytical prediction (\ref{eq:Wa}) quite accurately reproduces all key features of numerical spectra on Fig.~\ref{fig:spctr_alpha}, including the position of the main peaks and the structure of higher harmonics.

Of particular interest is the connection between the observed effects and systems where interactions within populations are dense but between populations are sparse. In the case of small subpopulations ($\alpha\ll 1$), their collective dynamics are determined less by the mean-field interaction with the full network and more by the individual dynamics of the elements within the subpopulation.

\section{Discussion}

In this work, we have investigated the shot noise generated by subpopulations within densely coupled neural populations. Understanding the spectral properties of these fluctuations is essential for accurate modeling of large-scale brain networks, where inter-population connections are typically sparse and involve only small, randomly selected subsets of neurons. As we have demonstrated, the signal transmitted through such sparse connections is inherently noisy, and its statistical characteristics depend critically on the size of the subpopulation. This dependence cannot be ignored when constructing stochastic mean-field models of hierarchical neural networks.

Our primary result is the analytical expression for the power spectral density of subpopulation shot noise, given by Eq. (23): \( W_\alpha = \alpha W + (1-\alpha)\tilde{W} \). This formula reveals that the spectrum of a subpopulation is not simply a scaled version of the full population spectrum. Instead, it arises from a nontrivial mixture of two distinct spectral components: the full shot noise spectrum \( W \) of the entire population and the free shot noise spectrum \( \tilde{W} \) that characterizes the intrinsic fluctuations of independent neurons. The mixing coefficient \( \alpha \)—the fraction of neurons belonging to the subpopulation—determines the relative contribution of each component. As the subpopulation becomes smaller (\( \alpha \to 0 \)), the spectrum approaches that of free shot noise, while for \( \alpha = 1 \), it reduces to the full population spectrum.

The validity of our analytical prediction has been thoroughly verified through direct numerical simulations, which show excellent qualitative and quantitative agreement across a wide range of subpopulation sizes (Fig. 3). The theory accurately reproduces all key features of the numerical spectra, including the position of the main low-frequency peak, the emergence and growth of the peak near the first harmonic of the mean frequency as \( \alpha \) decreases, and the structure of higher harmonics. This close correspondence confirms the correctness of our theoretical approach and its ability to capture the essential physics of finite-size fluctuations in neural populations.

A distinctive feature of our study is the use of a realistic, non-Lorentzian distribution of neuronal bias currents [Eq. (4)], whose tails decay as \( \eta^{-4} \). Unlike the Lorentzian distribution commonly employed in previous studies for mathematical convenience, our distribution has finite moments of all orders, making it physically plausible and eliminating unphysical divergences. Importantly, we have shown that our reduction technique and the nesting method generalize naturally to such distributions, demonstrating that the approach is not limited to the analytically convenient Lorentzian case. This extension significantly enhances the biological relevance of our results and provides a framework applicable to a broader class of heterogeneity profiles encountered in real neural systems.

The nesting method, originally developed for analyzing shot noise in whole populations, has proven its usefulness and universality by being successfully generalized to the case of subpopulations. The nested configuration—in which a finite-size population is embedded within an infinite superpopulation that shares the same parameter distribution—provides a mathematically rigorous foundation for separating microscopic and macroscopic fluctuation components. By further dividing the finite population into two subpopulations and carefully accounting for their interactions through the mean field, we were able to derive closed-form expressions for the subpopulation noise spectrum. The elegance of this approach lies in its ability to handle the recursive nature of the problem: the output of each subpopulation depends on the combined input from both subpopulations, which in turn depends on their outputs. The nesting method resolves this circularity by leveraging the thermodynamic limit of the superpopulation as a reference.

From a practical perspective, our findings provide the foundation for a new class of stochastic mean-field models for hierarchical neural networks. In such networks, where populations are densely connected internally but only sparsely connected to one another, the correct treatment of inter-population communication requires accounting for the size-dependent spectral fingerprint of subpopulation shot noise. Previous approaches either ignored noise altogether or employed ad hoc noise terms with unrealistic spectral properties. Our work supplies the analytical tools necessary to incorporate noise with the correct frequency dependence, enabling more accurate predictions of how fluctuations propagate through multi-area brain networks and how they interact with intrinsic dynamics to produce emergent phenomena such as noise-induced oscillations, stochastic resonance, or variability in information transmission.

Several limitations of the present study should be acknowledged. Our analysis assumed that the coupling strength \( J \) is constant and negative (inhibitory), and we focused on parameter regimes where the neural mass model possesses a unique stable equilibrium. The generalization to excitatory coupling or to regimes with limit cycle oscillations remains an important direction for future work. Additionally, we considered only the case of instantaneous synapses (\( \tau \to 0 \)); finite synaptic rise and decay times may introduce additional filtering effects that could modify the noise spectrum, particularly at high frequencies. Finally, our subpopulations were formed by random sampling from the full population, ensuring identical distributions of bias currents. In biological systems, subpopulations projecting to different targets may have distinct parameter distributions; extending our framework to such heterogeneous subpopulations would be a valuable extension.

\section{Conclusion}

In this paper, we have developed a comprehensive analytical framework for characterizing the shot noise generated by subpopulations within densely coupled neural populations. By generalizing the nesting method and employing a realistic non-Lorentzian distribution of neuronal parameters, we derived an explicit expression for the power spectral density of subpopulation shot noise that shows excellent agreement with numerical simulations. Our results reveal that this spectrum depends non-trivially on subpopulation size, arising from a mixture of full population noise and free shot noise rather than a simple scaled version of either. This work provides the analytical foundation for a new class of stochastic mean-field models for hierarchical neural networks with sparse inter-population connectivity, enabling accurate incorporation of size-dependent noise spectra in future studies of large-scale brain dynamics.

\begin{acknowledgments}
The work is supported by the Russian Science Foundation grant No. 25-22-00660.
\end{acknowledgments}

\section*{Data Availability Statement}

The data that support the findings of this study are available from the corresponding author upon reasonable request.

\section*{Appendix A. Sampling of the local parameters $\eta_j$}

We draw the bias currents of individual neurons of the population \eqref{a1_01} from distribution \eqref{eq:g} randomly and independently by the inverse transform method. For this sake, for each neuron we generate a random numbers $x_j$ from a uniform distribution $x\in[0,1]$ and determine $\eta_j$ by solving a nonlinear equation
\begin{equation}\label{a1_02_v4}
G(\eta_j)=x_j.
\end{equation}
Here $G(\eta)$ is the cumulative distribution function
\begin{equation}\label{a1_02_v5}
\begin{aligned}
G(\eta)=&\int_{-\infty}^{\eta} g(\eta')d\eta'=\frac{1}{2}+\frac{1}{4\pi}\ln\bigg\{\frac{\eta^2+\sqrt{2}\eta+1}{\eta^2-\sqrt{2}\eta+1}\bigg\}+\\
    &\frac{1}{2\pi}\big\{\arctan(\sqrt{2}\eta+1)+\arctan(\sqrt{2}\eta-1)\big\}.
\end{aligned}
\end{equation}

\section*{Appendix B. Derivation of the neural mass model}

In the limiting case $N \to \infty$, the state of population \eqref{a1_01} can then be described by the density function $\rho(V|\eta,t)$, which determines the fraction of neurons the membrane potential close to $V$ and the bias current close to $\eta$ at time $t$. The density $\rho$ obeys the  continuity equation
\begin{equation}\label{a1_03}
\partial_t \rho+\partial_V [(V^2+\eta+Js+I)\rho] = 0.
\end{equation}
It is easy to show that the stationary distribution $\rho_0(V|\eta) \propto (V^2+\eta+Js)^{-1}$ has the form of a Lorentzian function. Further we assume that the density $\rho(V|\eta,t)$ retains the Lorentzian form even in the nonstationary case and can be represented as
\begin{equation}\label{a2_01}
\rho(V|\eta,t)=\frac{1}{\pi}\frac{x(\eta,t)}{[V-y(\eta,t)]^2+x(\eta,t)^2},
\end{equation}
where the half-width $x(\eta,t)$ and the center $y(\eta,t)$ are  functions of time. It is easy to show that these functions are closely related to the mean firing rate and the mean membrane potential of the neurons with the bias current close to $\eta$:
\begin{eqnarray}
r(\eta,t)&=&\lim_{V\to\inf}=\rho(V|\eta,t)\dot{V}(V|\eta,t)=\frac{x(\eta,t)}{\pi},\\
v(\eta,t)&=&\text{P.V.}\int_{-\infty}^{\infty}\rho(V|\eta,t)V dV = y(\eta,t)
\end{eqnarray}\label{a2_02}
where P.V. denotes the Cauchy principal value. The entire population's mean firing rate $r(t)$ and mean membrane potential $v(t)$ are  given by
\begin{equation}\label{a2_04}
r(t)=\frac{1}{\pi}\int_{-\infty}^{\infty}x(\eta,t)g(\eta)d\eta,
\end{equation}
\begin{equation}\label{a2_05}
v(t)=\int_{-\infty}^{\infty}y(\eta,t)g(\eta)d\eta.
\end{equation}

The equations for the unknown functions $x(\eta,t)$ and $y(\eta,t)$ can be obtained by substituting (\ref{a2_01}) into (\ref{a1_03}) and separating the resulting expression into independent parts. Switching to a new variable $w(\eta,t)=x(\eta,t)+i y(\eta,t)$ allows the resulting equations to be expressed in complex form:
\begin{equation}\label{a2_06}
\partial_t w(\eta,t)=i[\eta+Js(t)-w(\eta,t)^2+I(t)].
\end{equation}
Note that in this case, the mean firing rate and mean membrane potential are related by:
\begin{equation}\label{a2_06a}
\pi r(t)+i v(t)=\int_{-\infty}^{\infty}w(\eta,t)g(\eta)d\eta.
\end{equation}

To close (\ref{a2_06}), the synaptic activation function (\ref{a1_02}) must be expressed in the new variables. For the case  $N\to\infty$ and $\tau\rightarrow 0$, the synaptic current $s(t)$ transforms into the mean firing rate $r(t)$, and we obtain

\begin{equation}\label{a2_08}
s(t)=r(t)=\frac{1}{\pi}\operatorname{Re}\int_{-\infty}^{\infty}w(\eta,t)g(\eta)d\eta.
\end{equation}

\begin{figure}
	\includegraphics[width=0.5\textwidth]{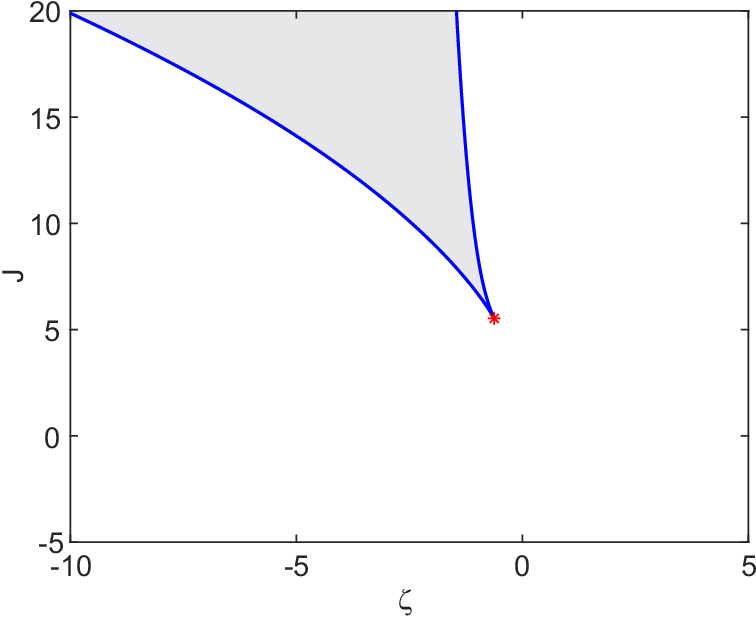}
	\caption{The bifurcation diagram of the neural mass model (\ref{a3_11}). The blue lines correspond to saddle-node bifurcations. The gray region highlights the bistability domain. A red asterisk marks the codimension-two bifurcation (cusp point). }\label{fig:BD_work}
\end{figure}

For rational functions $g(\eta)$, the integral in the r.h.s. of \eqref{a2_08} can be evaluated by the residue theorem. by closing the integration countour in the lower complex half-plane \cite{klinshov2021reduction}. If $w(\eta,t)$ is analytical and $g(\eta)$ has $n$ simple poles $\bar{\eta}_j$ in the lower complex half-plane, it is convenient to  define $h(\eta)=1/g(\eta)$ and write
\begin{equation}\label{a3_04}
\int_{-\infty}^{\infty}\frac{w(\eta,t)}{h(\eta)}d\eta=-2\pi i\sum_{j=1}^n {\operatorname{Res}_{\eta=\eta_j}} \frac{w(\eta,t)}{h(\eta)}=-2\pi i\sum_{j=1}^n \frac{w(\bar{\eta}_j,t)}{h'(\bar{\eta}_j)}.
\end{equation}
The distribution \eqref{eq:g} has two simple poles in the lower complex halfplane:
\begin{equation}\label{a3_06}
\begin{aligned}
   &\bar{\eta}_1=\zeta+\frac{\sqrt{2}}{2}-i\frac{\sqrt{2}}{2},\\
   &\bar{\eta}_2=\zeta-\frac{\sqrt{2}}{2}-i\frac{\sqrt{2}}{2}.
\end{aligned}
\end{equation}
whiche leads to
\begin{equation}
    r(t)=\frac{1}{2\pi}\operatorname{Re}\Big\{(1+i)w(\bar{\eta}_1,t)+(1-i)w(\bar{\eta}_2,t)\Big\}.
\end{equation}
By introducing $w(\bar{\eta}_1,t)=w_1(t)$, $w(\bar{\eta}_2,t)=w_2(t)$ and writing \eqref{a2_06} for $\eta=\bar{\eta}_1$ and $\eta=\bar{\eta}_2$ one obtains
\begin{equation}\label{a3_08}
\begin{aligned}
&\frac{dw_1(\bar{\eta}_1,t)}{dt}=i[\bar{\eta}_1+Jr(t)-w_1(\bar{\eta}_1,t)^2+I(t)],\\
&\frac{dw_2(\bar{\eta}_2,t)}{dt}=i[\bar{\eta}_2+Jr(t)-w_2(\bar{\eta}_2,t)^2+I(t)],\\
&r(t)=\frac{1}{2\pi}\operatorname{Re}\Big\{(1+i)w_1+(1-i)w_2\Big\}.
\end{aligned}
\end{equation}
By introducing new real variables $p$ and $q$ such that
\begin{equation}\label{a3_10}
\begin{aligned}
&w_1=p_1+i q_1,\\
&w_2=p_2+i q_2,
\end{aligned}
\end{equation}
one readily the neural mass model  \eqref{a3_11}.
The numerically computed bifurcation diagram of this model on the $J,\zeta$ parameter plane is presented in Figure~\ref{fig:BD_work}.
The system may possess up to four equilibrium states depending on parameters, though only one or two can be simultaneously stable. Notably, the system does not exhibit any limit cycles. When parameters are varied, the stable equilibrium states can undergo saddle-node bifurcations.  The corresponding lines in the bifurcation diagram bound a wedge-shaped region that corresponds to the bistability domain of the system.

\section*{Appendix C. Power spectrum of the free shot noise}

The distribution $g(\eta)$ can be represented as
\begin{equation}\label{a4_01}
g(\eta)=\frac{1}{N}\sum_{j=1}^N\delta(\eta-\eta_j).
\end{equation}
The distribution of natural frequencies can be written as follows
\begin{equation}\label{a4_02}
g_{\nu}(\nu)=\frac{1}{N}\sum_{j=1}^N\delta(\nu-\nu_j),
\end{equation}
where
\begin{equation}\label{a4_03}
\begin{aligned}
&\nu_j=\sqrt{\eta_j+Jr_0}/\pi, \text{    } &\eta_j+Jr_0>0,\\
&\nu_j=0, \text{    } &\eta_j+Jr_0\leq 0.
\end{aligned}
\end{equation}

The output signal of a single neuron $s_j(t)$ is a sequence of $\delta$-pulses with period $T_j=1/\nu_j$:
\begin{equation}\label{a4_05}
s_j(t)=\sum_{p=-\infty}^\infty \delta(t-p T_j - \Theta_j),
\end{equation}
where $p\in \mathbb{Z}$ determines the spike number in the sequence, and the individual phase shift of a single element is $\Theta_j\in [0;\pi)$.

The total output signal of the entire network then takes the form
\begin{equation}\label{a4_06}
s(t)=\frac{1}{N}\sum_{j=1}^N s_j(t).
\end{equation}

Then the autocorrelation function of this signal can be found as
\begin{equation}\label{a4_07}
\begin{aligned}
K(\tau)=\langle s(t)s(t+\tau)\rangle =\\
=\lim_{T\rightarrow \infty}\frac{1}{T}\int_0^T dt \frac{1}{N^2}\sum_{j=1}^N\sum_{p=-\infty}^{\infty}\delta(t-p T_j-\Theta_j)\times\\
\times\sum_{k=1}^N\sum_{q=-\infty}^{\infty}\delta(t-qT_k-\Theta_k+\tau).
\end{aligned}
\end{equation}

In this autocorrelation function, two components can be distinguished: $K(\tau)=K_1(\tau)+K_2(\tau)$, where $K_1(\tau)$ corresponds to the case $j=k$, and $K_2(\tau)$ corresponds to the case $j\neq k$.

Then
\begin{equation}\label{a4_08}
\begin{aligned}
K_1(\tau)=\lim_{T\rightarrow\infty}\frac{1}{T}\int_0^T dt \frac{1}{N^2}\sum_{j=1}^N \sum_{p=-\infty}^{p=\infty}\delta(t-pT_j-\Theta_j)\times\\
\times\sum_{q=-\infty}^{\infty}\delta(t-qT_j-\Theta_j+\tau)=\\
=\frac{1}{N^2}\sum_{j=1}^{N}\lim_{T\rightarrow\infty}\frac{1}{T}\int_0^T dt\sum_p\sum_q\delta(t-pT_j-\Theta_j)\times\\
\times\delta(t-qT_j-\Theta_j+\tau).
\end{aligned}
\end{equation}
Using the sifting property of the $\delta$-function, we compute this integral, setting $t=pT_j+\Theta_j$. Taking into account $0\leq t \leq T$, we obtain constraints on the possible values of $p$: $-\Theta_j/T_j\leq p \leq (T-\Theta_j)/T_j$, where $p$ is an integer.

As a result, we obtain
\begin{equation}\label{a4_09}
\begin{aligned}
K_1(\tau)=\frac{1}{N^2}\sum_{j=1}^{N}\lim_{T\rightarrow\infty}\frac{1}{T}\times\\
\times\sum_{\substack{-\Theta_j/T_j\leq p \leq (T-\Theta_j)/T_j, \\ p\in \mathbb{Z}}}\sum_{q=-\infty}^{\infty}\delta\big((p-q)T_j+\tau\big).
\end{aligned}
\end{equation}

Due to the unbounded limits over the parameter $q$, this function is insensitive to the shift caused by the change of the parameter $p$. Then
\begin{equation}\label{a4_10}
\begin{aligned}
K_1(\tau)=\frac{1}{N^2}\sum_{j=1}^N\lim_{T\rightarrow\infty}\frac{1}{T}\frac{T}{T_j}\sum_q \delta(-q T_j + \tau)=\\
=\frac{1}{N^2}\sum_j\frac{1}{T_j}\sum_q \delta(\tau-qT_j).
\end{aligned}
\end{equation}

Now we find the spectrum of this component of the autocorrelation function
\begin{equation}\label{a4_11}
\begin{aligned}
W_1(\omega)=\int_{-\infty}^{\infty} K_1(\tau) e^{-i\omega\tau}d\tau=\\
=\frac{1}{N^2}\sum_j \frac{1}{T_j}\sum_q \int_{-\infty}^{\infty}\delta(\tau-qT_j)e^{-i\omega\tau}d\tau=\\
=\frac{1}{N^2}\sum_j \frac{1}{T_j}\sum_q e^{-i\omega q T_j}.
\end{aligned}
\end{equation}

We further transform this expression using the periodicity property of the Dirac comb:
\begin{equation}\label{a4_12}
\begin{aligned}
W_1(\omega)=\frac{1}{N^2}\sum_j \frac{1}{T_j} \sum_q e^{-i\omega q T_j}=\\
=\frac{1}{N^2}\sum_{n=-\infty}^{\infty} \delta\bigg(\frac{\omega T_j^2}{2\pi}-n T_j\bigg)=\\
=\frac{1}{N^2}\sum_j \frac{1}{T_j^2}\sum_{\substack{n=-\infty, \\ n\in \mathbb{Z}}}^{\infty}\delta\bigg(\frac{\omega}{2\pi}-\frac{n}{T_j}\bigg).
\end{aligned}
\end{equation}

Or, converting from cyclic (angular) frequency directly to oscillation frequency, we obtain
\begin{equation}\label{a4_13}
W_1(\nu)=\frac{1}{N^2}\sum_j \nu_j^2 \sum_{\substack{n=-\infty, \\ n\in \mathbb{Z}}}^{\infty}\delta(\nu-n \nu_j),
\end{equation}
where it must be taken into account that $\nu \geq 0$.

Although the obtained expression completely describes the frequency spectrum of the function $K_1(\tau)$, it is still not very convenient for practical use. To correct this, we perform a series of further transformations, using the sifting property of the $\delta$-function and formula (\ref{a4_02}):
\begin{equation}\label{a4_14}
\begin{aligned}
W_1(\nu)=\frac{1}{N^2}\sum_{j=1}^N \nu_j^2 \sum_{\substack{n=-\infty, \\ n\in \mathbb{Z}}}^{\infty}\delta(\nu-n \nu_j)=\\
=\frac{1}{N}\sum_{n=-\infty}^{\infty}\int_{-\infty}^{\infty}df \frac{1}{N}\sum_{j=1}^N\delta(f-\nu_j)\delta(\nu-n f)f^2=\\
=\frac{1}{N}\sum_{n=-\infty}^{\infty}\int_{-\infty}^{\infty}df g(f)\delta(\nu-n f)f^2=\\
=\frac{1}{N}\sum_{n=-\infty}^{\infty}\int_{-\infty}^{\infty}df g(f) f^2\frac{1}{n}\delta\bigg(\frac{\nu}{n}-f\bigg)=\\
=\frac{1}{N}\sum_{\substack{n=-\infty, \\ n\in \mathbb{Z}}}^{\infty}\frac{1}{n} g\bigg(\frac{\nu}{n}\bigg)\frac{\nu^2}{n^2}.
\end{aligned}
\end{equation}

As a result, we obtain
\begin{equation}\label{a4_15}
W_1(\nu)=\frac{1}{N}\sum_{\substack{n=-\infty, \\ n\in \mathbb{Z}}}^{\infty}\frac{\nu^2}{n^3}g\bigg(\frac{\nu}{n}\bigg),
\end{equation}
where the values of $n$ must be chosen such that $\nu\geq 0$ holds. This corresponds to values $n\geq 0$, with $n=0$ corresponding to the frequency $\nu=0$. With this in mind, (\ref{a4_15}) can be rewritten as
\begin{equation}\label{a4_15a}
W_1(\nu)=\frac{1}{N}\sum_{\substack{n=0, \\ n\in \mathbb{Z}}}^{\infty}\frac{\nu^2}{n^3}g\bigg(\frac{\nu}{n}\bigg),
\end{equation}

Let us now determine the contribution to the spectrum of the total autocorrelation function $K(\tau)$ from its second component $K_2(\tau)$. In this case $j\neq k$ and
\begin{equation}\label{a4_16}
\begin{aligned}
K_2(\tau)=\frac{1}{N^2}\lim_{T\rightarrow\infty}\frac{1}{T}\int_{0}^{T}dt\times\\
\times\sum_{j}\sum_{k\neq j}\sum_p\sum_q\delta(t-pT_j-\Theta_j)\delta(t-qT_k-\Theta_k+\tau)=\\
=\frac{1}{N^2}\lim_{T\rightarrow\infty}\frac{1}{T}\times\\
\times\sum_{j}\sum_{k\neq j} \sum_{\substack{-\frac{\Theta_j}{T_j}\leq p \leq \frac{T-\Theta_j}{T_j}, \\ p\in \mathbb{Z}}}\sum_q\delta(pT_j+\Theta_j-qT_k-\Theta_k+\tau).
\end{aligned}
\end{equation}

Note that due to the incommensurability of $T_j$ and $T_k$, this expression is equivalent to the following:
\begin{equation}\label{a4_17}
K_2(\tau)=\frac{1}{N^2}\lim_{T\rightarrow\infty}\frac{1}{T}\sum_j\sum_{k\neq j}\sum_{l=0}^{\lfloor T/T_j \rfloor}\sum_{q=-\infty}^{\infty}\delta(\tau-q T_k+\Theta_l),
\end{equation}
where $\Theta_l=\Theta_l(p,\Theta_j,\Theta_k)$, $l\in\mathbb{Z}$, and the function $\lfloor \cdot \rfloor$ denotes taking the integer part of the argument. The values $\Theta_l$ are distributed over the half-interval $\Theta_l\in[0;T_k)$ irregularly, but on average uniformly. The average interval between neighboring $\delta$-pulses is:
\begin{equation}\label{a4_18}
\delta t =\lim_{T\rightarrow\infty}\frac{T_j T_k}{T}\rightarrow 0.
\end{equation}
Let us now introduce an auxiliary time interval $\Delta t$ such that $\delta t \ll \Delta t \ll T_j,T_k$. Next, we perform smoothing of the function $K_2(\tau)$ on this scale:
\begin{equation}\label{a4_19}
\begin{aligned}
K_2(\tau)=\frac{1}{\Delta t} \int_{\tau}^{\tau+\Delta t} K_2(\tau)d\tau=\frac{1}{N^2}\lim_{T\rightarrow\infty}\frac{1}{T}\times\\
\times\sum_j\sum_{k\neq j}\frac{1}{\Delta t}\int_{\tau}^{\tau+\Delta t}\sum_{l=0}^{\lfloor T/T_j \rfloor}\sum_{q=-\infty}^{\infty}\delta(\tau-q T_k+\Theta_l) d\tau=\\
=\frac{1}{N^2}\lim_{T\rightarrow\infty}\frac{1}{T}\sum_j\sum_{k\neq j}\frac{1}{\Delta t}\frac{\Delta t}{\delta t}=\lim_{T\rightarrow\infty}\frac{1}{T}\frac{T}{T_j T_k}=\frac{1}{T_j T_k}.
\end{aligned}
\end{equation}

Since we are interested in relatively slow fluctuations in the system, whose characteristic period is comparable to $T_j$ and $T_k$, the procedure of smoothing much faster fluctuations does not affect the final result.

Next, we convert in (\ref{a4_19}) from individual periods to natural frequencies:
\begin{equation}\label{a4_20}
\begin{aligned}
K_2(\tau)=\frac{1}{N^2}\sum_j\sum_{k\neq j}\frac{1}{T_j T_k}=\frac{1}{N^2}\sum_j\sum_{k\neq j}\nu_j\nu_k=\\
=\frac{1}{N^2}\Bigg[\sum_{j=1}^N\sum_{k=1}^{N}\nu_j\nu_k-\sum_{j=1}^N\nu_j^2\Bigg]=\langle \nu \rangle^2-\frac{1}{N}\langle \nu^2 \rangle.
\end{aligned}
\end{equation}

Now we find the power spectrum of this autocorrelation function
\begin{equation}\label{a4_21}
\begin{aligned}
W_2(\omega)=\int_{-\infty}^{\infty}K_2(\tau)e^{-i\omega\tau}d\tau=\\
=\bigg[\langle \nu \rangle^2-\frac{1}{N}\langle \nu^2 \rangle\bigg]\int_{-\infty}^{\infty}e^{-i\omega\tau}d\tau.
\end{aligned}
\end{equation}

Taking into account the property
\begin{equation}\label{a4_22}
\int_{-\infty}^{\infty}e^{-i\omega\tau}d\tau=-2\pi\delta(\omega),
\end{equation}
and converting from angular frequency directly to oscillation frequency, we obtain
\begin{equation}\label{a4_23}
W_2(\nu)=\bigg[\langle \nu \rangle^2-\frac{1}{N}\langle \nu^2 \rangle\bigg]\delta(\nu).
\end{equation}

As a result, for the total spectrum of the full autocorrelation function $K(\tau)$ we obtain
\begin{equation}\label{a4_24}
\begin{aligned}
W(\nu)=W_1(\nu)+W_2(\nu)=\\
=\frac{1}{N}\sum_{\substack{n=0, \\ n\in \mathbb{Z}}}^{\infty}\frac{\nu^2}{n^3}g\bigg(\frac{\nu}{n}\bigg)+\bigg[\langle \nu \rangle^2-\frac{1}{N}\langle \nu^2 \rangle\bigg]\delta(\nu)=\\
=\frac{1}{N}\langle\nu^2\rangle\delta(\nu)+\frac{1}{N}\sum_{\substack{n=1, \\ n\in \mathbb{Z}}}^{\infty}\frac{\nu^2}{n^3}g\bigg(\frac{\nu}{n}\bigg)+\\
+\bigg[\langle \nu \rangle^2-\frac{1}{N}\langle \nu^2 \rangle\bigg]\delta(\nu)=\langle \nu \rangle^2\delta(\nu)+\frac{1}{N}\sum_{\substack{n=1, \\ n\in \mathbb{Z}}}^{\infty}\frac{\nu^2}{n^3}g\bigg(\frac{\nu}{n}\bigg).
\end{aligned}
\end{equation}

\section*{Appendix D. Frequency response of the neural mass model}

Let us find the complex transfer coefficient for the mean-field model (\ref{a3_11}) describing the collective dynamics of the neural network. For this purpose, consider the effect of a weak external harmonic signal on the system
\begin{equation}
I(t) = \varepsilon e^{i\omega t}, \quad |\varepsilon| \ll 1,
\end{equation}
where $\varepsilon$ is the signal amplitude, and $\omega=2\pi\nu$ is its cyclic frequency.
In the linear approximation, the system oscillates near the equilibrium state $p_1=p_{10}$, $q_1=q_{10}$, $p_2=p_{20}$, $q_2=q_{20}$.
We seek the solution of system (\ref{a3_11}) in the form of small perturbations relative to this equilibrium state:
\begin{align*}
p_1 &= p_{10} + \varepsilon p_{11} e^{i\omega t}, \\
q_1 &= q_{10} + \varepsilon q_{11} e^{i\omega t}, \\
p_2 &= p_{20} + \varepsilon p_{21} e^{i\omega t}, \\
q_2 &= q_{20} + \varepsilon q_{21} e^{i\omega t},
\end{align*}
where $p_{k,1}$, $q_{k,1}$ ($k=1,2$) are normalized perturbation amplitudes of order $O(1)$. Substituting this solution into (\ref{a3_11}) and linearizing the right-hand sides, we obtain a system of equations
\begin{equation}
\begin{aligned}
i\omega p_{11}=&2q_{10}p_{11}+2p_{10}q_{11}\\
i\omega q_{11}=&\bigg(\frac{J}{2\pi}-2p_{10}\bigg)p_{11}-\bigg(\frac{J}{2\pi}-2q_{10}\bigg)q_{11}+\\
&\frac{J}{2\pi}p_{21}+\frac{J}{2\pi}q_{21}+1\\
i\omega p_{21}=&2q_{20}p_{21}+2p_{20}q_{21}\\
i\omega q_{21}=&\frac{J}{2\pi}p_{11}-\frac{J}{2\pi}q_{11}+\bigg(\frac{J}{2\pi}-2p_{20}\bigg)p_{21}+\\
&\bigg(\frac{J}{2\pi}+2q_{20}\bigg)q_{21}+1
\end{aligned},
\end{equation}
Solving this system by substitution (and assuming $p_{10}\neq0$ and $p_{20}\neq 0$), we find:
\begin{equation}
    q_{11}=\frac{p_{11}(i \omega-2q_{10})}{2p_{10}},
\end{equation}

\begin{equation}
    q_{21}=\frac{p_{21}(i \omega-2q_{20})}{2p_{20}},
\end{equation}
\begin{equation}
    p_{21}=\frac{p_{20}\big[Jp_{11}\big(i\omega-2(p_{10}+q_{10})\big)-4\pi p_{10}\big]}{p_{10}\big[J\big(i\omega+2(p_{20}-q_{20})\big)+2\pi P\big]},
\end{equation}
\begin{equation}
    p_{11}=\frac{2\pi p_{10} P}{Q},
\end{equation}
where the auxiliary functions $P$ and $Q$ have the form
\begin{equation}
    P=\omega^2+4 i q_{20}\omega-4R_2^2,
\end{equation}
and
\begin{equation}
\begin{aligned}
    &Q=-\pi[\omega^2+4 i q_{10}\omega-4R_1^2][\omega^2+4 i q_{20}\omega-4R_2^2]-\\
    &J\big[(p_{10}-q_{10}+p_{20}+q_{20})\omega^2-2i\big(R_1^2-R_2^2-\\
    &2(p_{20}q_{10}+p_{10}q_{20})\big)\omega-4\big((p_{10}+q_{10})R_2^2+(p_{20}-q_{20})R_1^2\big)\big],
\end{aligned}
\end{equation}
with auxiliary notations $R_1^2=p_{10}^2+q_{10}^2$, $R_2^2=p_{20}^2+q_{20}^2$.

The complex transfer coefficient $S(\nu)$ is defined as the ratio of the amplitude of the output signal $r(t)$ (the network's mean frequency) to the amplitude of the input signal $I(t)$. As a result, we obtain
\begin{equation}
    S(\nu)=\frac{1}{2\pi}(p_{11}-q_{11}+p_{21}+q_{21}).
\end{equation}

\section*{references}

\end{document}